\documentclass[epj,referee]{svjour}
\usepackage{graphics}
\begin{document}

\title{Effects of Saving and Spending Patterns on Holding Time Distribution}
\author{Ning Ding \and Ning Xi \and Yougui Wang 
\thanks{\emph{E-mail:} ygwang@bnu.edu.cn }
}                     
\offprints{Yougui Wang}          
\institute{Department of Systems Science, School of Management,
Beijing Normal University, Beijing, 100875, P.R. China}
\date{Received: date / Revised version: date}
%
\abstract{The effects of saving and spending patterns on holding
time distribution of money are investigated based on the ideal
gas-like models. We show the steady-state distribution obeys an
exponential law when the saving factor is set uniformly, and a
power law when the saving factor is set diversely. The power
distribution can also be obtained by proposing a new model where
the preferential spending behavior is considered. The association
of the distribution with the probability of money to be exchanged
has also been discussed. 
} 
\maketitle
\section{Introduction}\label{intro}

During the last several years, methods and techniques of
statistical physics have been successfully applied to economical
and financial problems \cite{ephy1,ephy2,ephy3}. Recently, Some
econophysists have been paying attention to the statistical
mechanics of money, theoretically or empirically
\cite{redner,basic,saving1,bkccom,saving2,saving3,emp1,emp2,emp3}.
They believe that a thorough understanding of the statistical
mechanics of the money, especially studying of the distribution
functions, is essential. Some pioneering work along this line has
been reviewed in a popular article \cite{hayes}.

As well known, the exploration of the distribution of money can be
traced back at least a century to the work of the Italian social
economist  Vilfredo Pareto, who studied the distribution of income
among people in different western countries and found an inverse
power law \cite{pareto}. Recently this topic has been taken up
with the emergence of econophysicists among whom some believe that
there might be some physical and mathematical rules governing the
distribution of income or wealth in the world and attempt to
discover them. A series of models have been developed for the
equilibrium money distribution based on the analogy between market
economics and kinetic theory of gases
\cite{redner,basic,saving1,saving2,emp3}. Identifying exchange
between any two agents in a closed economy where the money is
conserved with the two-body elastic collision in an ideal gas,
these models show no matter how uniformly and forcefully one
distributes money among agents initially, the successive tradings
eventually lead to a steady distribution of money. And the shape
of money distribution is determined by the trading rule for
choosing an amount of money to transfer. Allowing agents to hold
back some of their money when they are chosen to trade, B.K.
Chakraborti et al. introduced the saving behavior into the model
by adding a saving factor $s$ in the trading rule \cite{saving1}.
The simulation results clearly indicate a robust Gibbs-like
distribution where the density of agents with money $m$ decreases
exponentially with $m$ for $s=0$, which is identical to the result
of A. Dr\u{a}gulescu and V. M. Yakovenko's random two-agent
exchanges model \cite{basic}. The distribution of money changes to
follow asymmetric Gibbs-like law when the fixed and uniform saving
factor is set to be nonzero, while  a `critical' Pareto
distribution of money is found when saving factor is set diversely
among agents \cite{saving1,bkccom,saving2}.

In practice, money is held as a store of value, what is more, it
plays an essential role for being a medium of exchange. Money is
transferred consecutively from hand to hand in the exchange
process, in which there exist time intervals for money to be held.
This kind of time interval was ever called by Wicksell the
``average period of idleness'' or ``interval of rest'' of money
\cite{Wicksell}. In our previous work \cite{circulation}, we
called it ``holding time'' of money and found that after the
economy has achieved an equilibrium state, there is not only a
distribution of money among agents, but also a steady distribution
over the holding time. We also found that monetary velocity, an
important macroeconomic variable, which is associated with Irving
Fisher \cite{fish}, could be expressed as the expectation of the
reciprocal of holding time.

In a basic ideal gas-like model, the distribution of money over
the holding time follows an exponential law, where saving behavior
is not taken into account. The purpose of this paper is to study
how the introduction of saving behavior affects such kind of
distribution. In next section, we make a brief review of the basic
ideal gas-like model by which our work can be erected and of the
measurement of the distribution of holding time. In sections
\ref{saving} and \ref{diff}, we show that the uniform saving
factor gives exponential distribution, while the diverse saving
factor induces a change to power distribution. Then we introduce
preferential spending behavior into the model in section
\ref{prefer} and again obtain power distribution. Comparing these
results, we can conclude that the formation of holding time
distribution is associated with the character of the probability
of money to be exchanged.

\section{An Ideal Gas-Like Market Model and Holding Time Distribution}\label{Basic}

We begin with the basic ideal gas-like model which was introduced
firstly by A. Dragulescu and V.M. Yakovenko \cite{basic}. A close
economy is considered in the model where the amount of money $M$
is conserved and the number of agents $N$ is fixed. The money is
possessed by agents individually and agents can exchange money
with each other. Since the scale and initial distribution of money
have no effect on the final results, most of our simulations were
carried out with $N=250$ and $M=25000$ and the amount of money
held by each agent was set to be $M/N$ at the beginning. The trade
in the economy is modelled to take place round by round. In each
round, two agents, $i$ and $j$ for example, are chosen randomly to
get engaged in a trade among which agent $i$ is "receiver" and the
other one $j$ is "payer" . The amount of money that changes hand
$\Delta m$ is determined by trading rule which ensures that the
amount of any agent's money is non-negative and the total money is
conserved. A trading rule commonly used can be expressed as
$\Delta m=\varepsilon(m_i+m_j)/2$, where $\varepsilon$ is a random
number from zero to unity. As for which units of money are chosen
to be transferred, all in the payer's hand is equally probable.

In the ideal gas-like model, money is held by agents and
transferred frequently. In this process, if an agent receives
money from other agents, he will hold it in hand till paying it
out to some other agents. The time interval between the receiving
and paying out is named as holding time \cite{circulation}. The
holding times of a certain unit of money at different moments or
those of different units of money at a given moment are not the
same. We introduce the probability distribution function of
holding time $P_h(\tau)$, which is defined so that the amount of
money whose holding time lies between $\tau$ and $\tau+d\tau$ is
equal to $MP_h(\tau)d\tau$. So, we can get the normalization
condition and the expression of the expectation of holding time as
follows:
\begin{equation}\label{normal}
    \int^\infty_0 P_h(\tau)d\tau=1
\end{equation}
and
\begin{equation}\label{expect}
    T=\int^\infty_0 \tau P_h(\tau)d\tau.
\end{equation}

In the simulations, suppose it is at round $t_0$ that we start to
reord, and so, holding time is recorded as the difference between
the moments when the money takes part in trade after $t_0$ for the
first two times. The recording mode is illustrated in Figure 1(a).
Please note this mode is different from what we have adopted in
Reference \cite{circulation}, which is shown in Figure 1(b). The
measurement results of the two modes seem quite different,
however, they reflect the same process in different ways. We adopt
herein the mode (a) solely to facilitate the exposition. The
typical distribution of holding time is shown in Figure 2. It can
been seen from the inset of Figure 2 that the distribution of
holding time follows an exponential law:
\begin{equation}\label{holdtime}
P_h(\tau)=\frac{1}{T}e^{-\frac{\tau}{T}}.
\end{equation}
This result indicates that the transferring process of money is a
Poisson process with intensity of $\frac{1}{T}$.

To get the distribution of holding time without systematic factor
disturbing, we performed the simulations about 100 times with
different random seeds and data were not collected until the
probability distribution of money got stationary. And for
convenience we stopped data collecting after majority of
money($>99.9\%$) had been recorded. In all the following
simulations, holding times are measured in this way, after the
distributions of money get stationary of course.

\section{Model with Uniform Saving Factor}\label{saving}

In reality, saving behavior is a natural action pattern for any
economic agent. In order to insure future consumption, people
always keep a part of their money as saving. The ratio of the
saving to total amount of money held by an agent is called
``marginal propensity to save'' by B. K. Chakrabarti's group. The
term of marginal propensity to save has totally different meaning
in economics, which is defined as the partial derivative of saving
function with respect to income \cite{econo}. To avoid confusion,
we rename it as ``saving factor''. Referring to the saving factor,
two cases have been considered by Chakrabarti's group, one is that
all agents have a uniform saving factor, the other one is that
saving factors are randomly distributed among agents
\cite{saving1,saving2}. As mentioned above, they found the
equilibrium distributions of money among agents had remarkable
different characters under such two assumptions. Along this line,
in this section and the next one we shall examine the impacts of
the saving behavior on the distribution of holding time for the
two cases respectively.

All the assumptions of the above ideal gas-like model do work in
this model. The amount of money is conserved and the number of
agents is fixed. Any agent's money is non-negative or no debt is
allowed. The agents are indistinctive at the beginning of
simulations: same initial amount of money and same saving factor
$s$. In each round, an arbitrary pair of agents are chosen to make
exchange with each other. For example, at $t$-th round, agent $i$
and $j$ take part in trading, so that at $t+1$-th round their
money $m_i(t)$ and $m_j(t)$ change to
\begin{equation}\label{aaa}
m_i(t+1)=m_i(t)+\Delta m; m_j(t+1)=m_j(t)-\Delta m;
\end{equation}
where \begin{equation}\label{deltam1}
    \Delta m=(1-s)[(\varepsilon-1)m_i(t)+\varepsilon
    m_j(t)];
\end{equation}
and $\varepsilon$ is a random fraction. After a straight-forward
substitution, it is obvious that the trading rule satisfies the
conservation and non-negativity condition, and each agent saves
fraction $s$ of his money before trade.

The simulation results are shown in Figure 3, for some values of
$s$. It can be seen that the probability distributions of holding
time for all saving factors decay exponentially. And the lower the
saving factor is, the steeper the distribution curve. These
results indicate this kind of saving behavior does not change the
Poisson nature of the exchanging process, but its intensity.

\section{Model with Diverse Saving Factor} \label{diff}
In realistic economy, how much an agent saves depends on the
economic situations he or she faces, and the saving factor of
course varies from agent to agent due to their different
conditions. To get closer to reality, this model inherits all the
assumptions and evolution mechanism of the previous model except
that of uniform saving factor. Each agent's saving factor is
initialized at the beginning of simulations which distributes
randomly and uniformly within an interval 0 to 1, and is fixed in
the simulations. Correspondingly, the trading rule Equation
(\ref{deltam1}) changes to
\begin{equation}\label{deltam2}
    \Delta m=(1-s_i)(\varepsilon-1)m_i(t)+(1-s_j)\varepsilon
    m_j(t);
\end{equation}
where $s_i$, $s_j$ are the saving factors of agent $i$ and $j$
respectively.

To our surprise, once the diverse saving factor is introduced into
the model, as shown in Figure 4, the holding time distribution
changes to obey a power law instead of an exponential law. This
result indicates that the transferring process of money in this
model is not a Poisson process any more. The Poisson nature of the
process is broken due to the loss of homogeneity of the money
transferring. In the previous model, for any saving factor, the
probability of each unit of money to participate in exchanges at
any round is equal because the saving factor is set to be uniform
for all agents. On the contrary, in this model the transferring
probability of money is not the same any more due to the diversity
of the saving factors. This conclusion was verified by two further
measurements on the exchange process.

Firstly, we measured the correlation coefficient of agents' saving
factors and the amount of money in their hands. As shown in Figure
5, the correlation coefficient increases sharply at the beginning
of simulation, starts to decrease slowly after about 2000th round.
The reason of the reduction is that the correlation coefficient
can not pick up non-linear associations. We also found the
correlation coefficient falls to and keeps at about 0.32 after
500000th round. Although the value of the correlation coefficient
is not high enough, it still implies that the agents with higher
saving factors hold more money.

Secondly, we computed the average value of saving factors over
total money corresponding to their respective holders after the
steady distribution of money among agents had been observed. The
value is 0.86 which certifies again that there is more money in
the hands of holders with higher saving factors. The average value
of saving factors over the money transferred was also computed,
its value is about 0.52. This fact says that the money held by
agents with higher saving factor has lower probability to take
part in trade. If all money has equal probability, combining with
the fact that the agents with higher saving factors hold more
money, it can be deduced that the value of this kind of average
saving factor should be about 0.86 all the time. Thus, we can
conclude that the higher the saving factor of a unit of money's
holder, the smaller probability for it to be transferred.

\section{Model with Preferential Spending} \label{prefer}

From the previous two models, we can see that different saving
patterns lead to different holding time distributions. Especially,
when the agents' saving factors are diverse, the probability of
money to take part in trade differs. Nevertheless, the
probabilities of the money held by the same one agent are equal to
each other. This is an implicit assumption in all the simulations
which means the money is homogenous to any agent. However, it is
not the case in real life. As the medium of exchange, money
changes hand to hand. In this circulation process, money abrades
unavoidably. And when agents make exchange, the payers might spend
their money with preference according to the degree of abrasion.
As a result, the money is not homogenous for agents. To overcome
this unrealistic feature, we proposed a new model which is quite
similar to the model with uniform saving factor. The only
alteration is that the probability of money chosen to change hand
is not equal even if the money is held by the same agent, in other
words, that the agents spend money with preference.

In each round, two agents, $i$ and $j$, are chosen randomly to
participate in the trade. The amount of money transferred is
determined by Equation (\ref{deltam1}). If agent $i$ is the payer,
the probability of money $k$ among $m_i$ to be transferred is
given by:
\begin{equation}\label{pre}
    p(k)=\frac{l_k+1}{\displaystyle \sum_{n=1}^{m_i}(l_n+1)};
\end{equation}
where $l_n$ is the times that money $n$ has participated in the
trade since the beginning of simulation. Here, we express the
probability with the sum of exchange times and 1 instead of
exchange times itself in case that denominator be zero at the
beginning of simulations.

The probability distributions of holding time for several
different saving factors are recorded after money distributions
reach stationary state which are shown in Figure 6. All
distributions obey power law, and the only difference is the
exponent.

The power distribution arises from the diversity of the
probability of money to participate in exchanges. At beginning of
our simulation, no money has ever taken part in the trade, thus
the probabilities are equal for all money according to Equation
(\ref{pre}). After some of money are exchanged randomly, they have
higher probabilities and the others have relative lower ones. As
the times of exchange increase, this slight diversity of money in
the probability will be enlarged till a stable distribution is
formed. To see this process from another point of view, the longer
for one unit of money to wait, the lower probability for it to be
spent. In this way, comparing with the case without preference,
some money's holding times get shorter, while some get much
longer. Thus the power distribution appears.

We studied the holding time distribution at different times, and
found the power distribution is robust. For instance, the holding
time distribution for $s=0$ still has the power form even after
$t=500000$. Contrarily to this, it is just after $t=1000$ that one
can clearly observe the steady distribution.

\section{Conclusions} \label{conc}

In this paper, the effects of saving and spending patterns on the
distribution of money over holding time are examined by computer
simulations. All the simulations are performed basing on the ideal
gas-like models. We consider two kinds of assumptions on saving
pattern, one is that all agents have uniform saving factor, the
other one is that the saving rates are set randomly distributed
among the agents. In the model with uniform saving factor, the
distribution of money over the holding time follows an exponential
law, while in the model with diverse saving factor the
distribution changes to a power type. We further propose a new
trading model where the agents spend money with preference and
also get power distribution. The simulation results indicate that
the final distribution is determined by the character of the
probability that money is chosen to participate in the trade.

\section*{Acknowledgments}
We would like to thank Zengru Di and Jinshan Wu for useful
comments and discussions. This work was supported by the National
Science Foundation of China under Grant No. 70071037.

\begin{figure}
\resizebox{0.45\textwidth}{!}{%
  \includegraphics{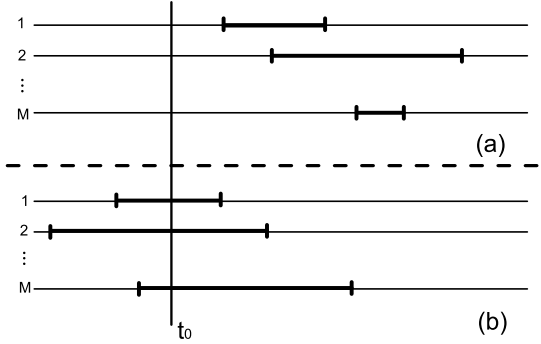}
} \caption{ Schematic presentation of the sampling method of
holding time adopted: (a) in this paper; (b) in Reference
\cite{circulation}. $t_0$ denotes the sampling time point, the
light horizontal solid lines represent the evolution history of
money, the vertical short bars symbolize moments for corresponding
money to be transferred and then the dark segments correspond to
the holding times to be recorded.} \label{fig:1}
\end{figure}

\begin{figure}
\resizebox{0.45\textwidth}{!}{%
  \includegraphics{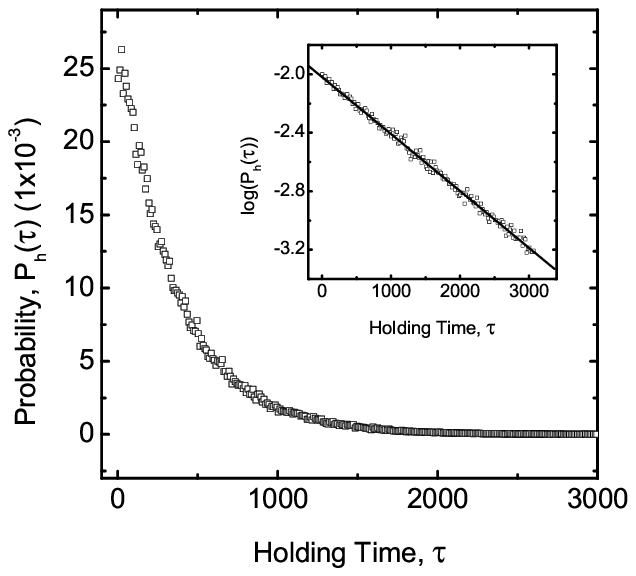}
}

\caption{The stationary holding time distribution obtained from
the basic ideal gas-like model simulations versus holding time.
The fitting in the inset indicates the distribution follows the
exponential law: $P_h(\tau)=\frac{1}{T}exp(-\tau/T)$.}
\label{fig:2}
\end{figure}

\begin{figure}
\resizebox{0.45\textwidth}{!}{%
  \includegraphics{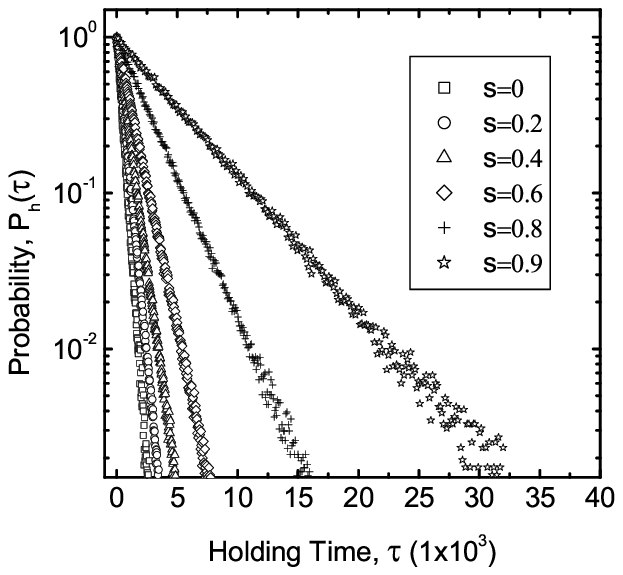}
} \caption{The stationary distributions of holding time for
several saving factors from 0 to 0.9 derived from the simulations
of the model with uniform saving factor in the semi-logarithmic
scale. Note that in the figure the probabilities have been scaled
by the maximum probability respectively.} \label{fig:3}
\end{figure}

\begin{figure}
\resizebox{0.45\textwidth}{!}{%
  \includegraphics{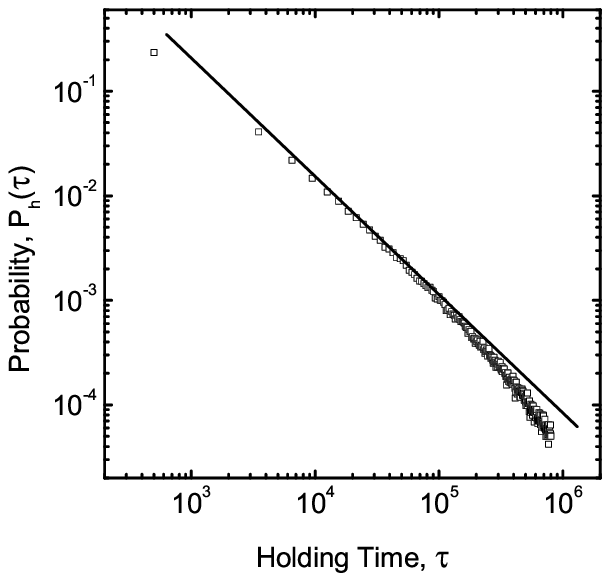}
} \caption{The stationary distribution of holding time derived
from the simulations of the model with diverse saving factor in
double logarithmic scale. The solid line is numerically fitted
line in the form of $P_h(\tau)\propto \tau^{-1.14}$.}
\label{fig:4}
\end{figure}

\begin{figure}
\resizebox{0.45\textwidth}{!}{%
  \includegraphics{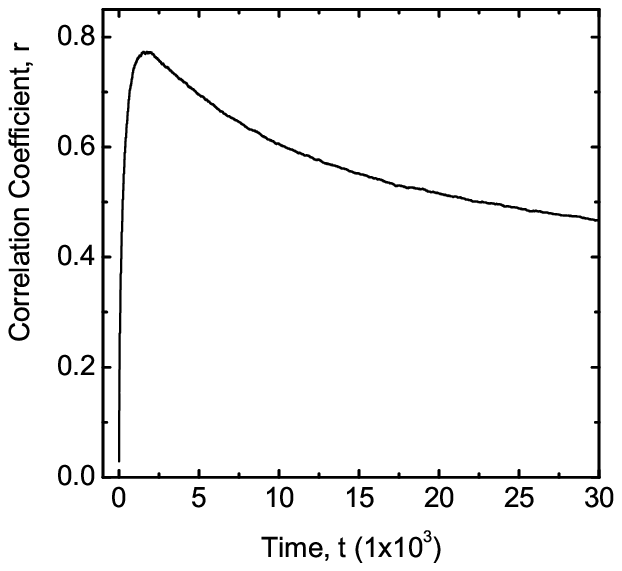}
} \caption{Correlative coefficient between the amount of money
held by agents and their saving factors versus time. At time
$t=1541$, the coefficient reaches its maximum 0.773.}
\label{fig:5}
\end{figure}

\begin{figure}
\resizebox{0.45\textwidth}{!}{%
  \includegraphics{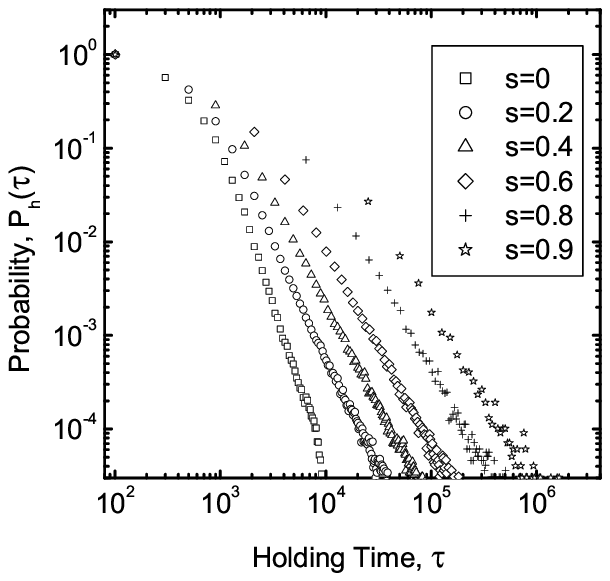}
} \caption{The stationary distributions of holding time for
several uniform saving factors from 0 to 0.9 derived from the
simulations of the model with preferential spending in the double
logarithmic scale. Note that in the figure the probabilities have
been scaled by the maximum probability respectively.}
\label{fig:6}
\end{figure}

\end{document}